\documentclass[12pt]{iopart}

\usepackage{graphicx}

\begin{document}

\title{Electrically charged gravastar configurations}

\author{Dubravko Horvat, Sa{\v{s}}a Iliji{\'c} and Anja Marunovi{\'c}}

\address{Department of Physics,
  Faculty of Electrical Engineering and Computing,
  University of Zagreb, Unska 3, HR-10\,000 Zagreb, Croatia}

\ead{dubravko.horvat@fer.hr, sasa.ilijic@fer.hr, anja.marunovic@fer.hr}

\begin{abstract}
The notion of a compact object immune to the horizon problem
and comprising an anisotropic inhomogeneous fluid
with a specific radial pressure behavior, i.e.\ the gravastar,
is extended by introducing an electrically charged component.
Einstein--Maxwell field equations are solved
in the asymptotically de~Sitter interior
where a source of the electric field is
coupled to the fluid energy density.
Two different solutions which satisfy the dominant energy condition are given:
one is the $\delta$-shell model
for which the analysis is carried out within Israel's thin shell formalism,
the other approach --- the continuous profile model ---
is solved numerically and the interior solutions
have been (smoothly) joined with the Reissner--Nordstr\"om exterior.
The effect of electric charge is considered,
and the equation of state, the speed of sound
and the surface redshift are calculated for both models.
\end{abstract}

\pacs{
04.20.-q  
04.40.-b  
04.40.Nr  
}

\section{Introduction}

The search for solutions of Einstein's equations
with different inputs for the physical content (dust, perfect fluid, etc.)
and features (charge, anisotropy, viscosity, etc.)
has been one of the steppingstones on the way to achieving
comprehensible picture of today's Universe.
Among solutions found so far one eminent position certainly belongs
to black hole solutions with intriguing properties and characteristics
(see for instance Ref.~\cite{AdBBLH} for a recent review).
However, some of the black hole features have been shown to be unattractive,
apart from the problem that irrefutable observational evidence
of the black hole existence has not yet been found.
These reasons had inspired the search for alternative configurations
which led to a solution dubbed \emph{gravastar},
the \emph{gra}vitational \emph{va}cuum \emph{star},
by Mazur and Mottola \cite{MazMot01,MazMotPNAS}.
These spherically symmetric static solutions to the Einstein equations
--- candidates for highly compact astrophysical objects
and in this sense alternatives to black holes ---
evolve from the segment of the de~Sitter geometry in the center
with the equation of state $p+\rho=0$,
proceed through a thin vacuum phase transition layer,
avoid the event horizon formation,
and swiftly match the exterior Schwarzschild spacetime.
(Objects with asymptotically de~Sitter core
were considered also in earlier literature,
see e.g.\ Refs~\cite{Gliner66,Sakharov66,Dym92}.)
One development of the gravastar idea went in the direction of
generalizing the interior and the exterior geometries
and modeling the phase transition layer by a $\delta$-shell
\cite{VissGS1,CarterStableGS,BilicGS,LoboStableDES,LoboNLED,DHSI07,Nilton08}.
Another direction of development focused on replacing
the $\delta$-shell by a continuous transition of the
energy density and pressure profiles from the
asymptotically de~Sitter center to the exterior Schwarzschild metric
\cite{VissGS2,DeBenedGS1}.
The common feature of all realizations of gravastars
is the anisotropy of principal pressures
\cite{BowLi74,HerreraSantos97,HerrOspPri08}.
Several astrophysically relevant aspects of gravastar solutions
such as thermodynamic properties \cite{BroderickNarayan07},
modes of quasi-normal oscillations \cite{ChirentiRezzolla07},
and ergoregion instability \cite{CardosoGS,ChirErgo08}
were also discussed in the literature.

Charged anisotropic models were considered
in a number of papers \cite{SSHelmi95,RoyRR96,KrishnaRAT00,HerrVar96}. 
Requiring $T_r^r=0$ for a charged anisotropic fluid,
the solutions may represent electromagnetic mass models
of electrically neutral systems \cite{Gron86},
since the total charge must be zero
if one wants to have a configuration
with the vanishing radial pressure at the surface.
Another charged object with neutral core
and the electric charge distributed on a $\delta$-shell
was investigated in Ref.~\cite{Varela07}.
In the same context the anisotropic charged spheres
with conformal motion/symmetry were considered
in Refs~\cite{KrishnaRAT00,HerreraPdL85}.
Anisotropic charged spheres with (varying) cosmological
constant were considered in Ref.~\cite{RayBhadra04}.

In this paper the gravastar picture is extended
to include the effect of the electric charge
as a natural step in gravastar investigations.
Although astrophysical objects are essentially neutral,
the problem of the electric charge in the phenomenological context
could occur in the (strange) quark stars considerations,
or accreting objects.
Also, the influence of the electric charge on the spacetime curvature
and other features of the Einstein--Maxwell system,
could be seen in the context of
the model of a classical charged massive particle.
Therefore the electric charge extension in the gravastar context
could be understood as a natural step in investigations.
The solutions for the charged gravastar obtained here
satisfy the dominant energy condition (DEC) everywhere
and possess no horizons.
The DEC requires the energy density to be non-negative,
and the absolute values of the pressures
not to exceed the value of the energy density.
It is taken as the principal criterion for the viability of solutions.
(The less stringent weak and null energy conditions
are automatically satisfied if the DEC is satisfied \cite{VisserBook}.)
The strong energy condition (SEC), which requires that
the sum of the energy density and all of the pressures is non-negative
(and guarantees the attractive character of gravity),
is by definition violated
in the de~Sitter core of the gravastar \cite{VissGS1,DHSI07}.
The interior metric is obtained by solving the Einstein-Maxwell equations
in which the matter --- anisotropic inhomogeneous charged fluid ---
through the Gauss law constraint of electrostatics
serves as a source for the electric field.
Such charged objects induce the Reissner-Nordstr\"om (RN) exterior metric
which smoothly joins the regular interior solution.

Two different charged gravastar models are presented.
In Sec.~\ref{sec:delta} we consider a solution with
a $\delta$-shell charge distribution.
We derive bounds on the configuration parameters
that follow from the requirement that the DEC is satisfied
and that the emergent equation of state
leads to subluminal speed of sound.
Physically viable solutions with unbounded surface redshift
are shown to exist for near-to-extremally charged $\delta$-shell gravastars.

Another solution presented in Sec.~\ref{sec:continous}
stems from the notion of the anisotropic fluid
with continuous pressure profiles
and with an electrically charged component
coupled to the fluid energy density
\cite{Gron86,PdL87,PdL93,HerrVar96,PdL04}.
This choice simplifies the calculations,
but does not restrict the overall generality
since the $\delta$-shell model of Sec.~\ref{sec:delta}
can be understood as the model of perfectly conducting sphere
with the electric charge expelled to the surface.
The numerically obtained solutions reveal a strong dependence
on the amount of the electric charge,
but not on the particular shape of the assumed fluid energy density profile
(as long as it obeys usual requirements \cite{BaumRend93}).
For both solutions with different charge content
the redshift is calculated,
the equation of state (EOS) is inferred in the atmosphere of the gravastar
where standard physics is expected to be valid \cite{VissGS2}.
Also in this region the speed of sound is calculated
to test the viability of results.
Conclusions are given in Sec.~\ref{sec:concl}.


\section{The $\delta$-shell model \label{sec:delta}}

\subsection{The model}

In the single $\delta$-shell picture of the electrically charged gravastar,
the interior geometry is a segment of the de~Sitter (dS) spacetime,
while the exterior is the asymptotically flat segment
of the Reissner--Nordstr\"om (RN) spacetime.
The two metrics are joined at a spherical timelike hypersurface
representing the `vacuum phase transition' layer
of the original Mazur--Mottola model.
Israel's thin shell formalism \cite{Israel66}
formulates the conditions for the smooth joining
of the metrics at a hypersurface.
The first junction condition
requires that the metric on the hypersurface
induced by the metrics on its two sides be the same.
The second junction condition requires that
the extrinsic curvature of the hypersurface
be the same when computed from the metrics on the two sides.
If the second junction condition is not satisfied,
then the joining is not smooth,
but it is still allowed by the Einstein equations
provided that a particular distribution of energy-momentum
is introduced on the hypersurface.
Such distribution is understood as $\delta$-shaped in the complete spacetime.

The dS metric of the gravastar interior
and the RN metric of the exterior spacetime
can both be written using the geometrized units
and the coordinates $x^{\alpha}=(t,r,\vartheta,\varphi)$ as
   \begin{equation}
   d s^2 = - k \, \big(1-\mu(r)\big) \, dt^2
           + \big(1-\mu(r)\big)^{-1} dr^2 + r^2 \, d\Omega^2 \; ,
   \label{eq:mumetric}
   \end{equation}
where $\mu(r)$ is the so-called compactness function,
$k$ is the time coordinate scaling constant
and $d\Omega^2$ is the metric on the unit sphere.
The compactness vanishes at the regular center of a spherical body,
and also at the asymptotically flat spatial infinity,
while for all $r$ it must remain less than unity
in order to avoid formation of horizons in the spacetime.
The junction hypersurface is specified with its radius which we denote $r=a$.
The metric on the junction hypersurface induced
from the general metric (\ref{eq:mumetric})
can be written adopting the coordinates $y^a=(t,\vartheta,\varphi)$ as
   \begin{equation}
   d \Sigma^2 = - k \, \big(1-\mu(a)\big) \, dt^2 + a^2 \, d\Omega^2 \; .
   \end{equation}
The first junction condition is automatically satisfied provided that
$ k_{\mathrm{dS}} ( 1 - \mu_{\mathrm{dS}}(a) )
   = k_{\mathrm{RN}} ( 1 - \mu_{\mathrm{RN}}(a) ) $,
where we choose to put $k_{\mathrm{RN}} = 1$ as usual.
The non-vanishing components
of the extrinsic curvature of the junction hypersurface,
computed from the general metric (\ref{eq:mumetric}), are
   \begin{equation}
   K_t^t = - \frac{\mu'(a)}{2\sqrt{1-\mu(a)}}
   \quad , \quad
   K_{\vartheta}^{\vartheta} = K_{\varphi}^{\varphi} =
   \frac{1}{a} \sqrt{1-\mu(a)}
   \label{eq:kab}
   \end{equation}
(prime denotes derivative with respect to $r$).
According to the thin shell formalism,
the hypersurface energy-momentum tensor is given by
   \begin{equation}
   S^a_b = - \frac{1}{8\pi}
             \Big( \big[K^a_b\big] - \big[K\big] \delta^a_b \Big) \; ,
   \label{eq:sab}
   \end{equation}
where $K=K_a^a$ and the square brackets indicate
the discontinuity of the quantity across the hypersurface,
in our case $[f] = \lim_{\epsilon\to0} f(a+\epsilon)-f(a-\epsilon)
= f_{\mathrm{RN}}(a) - f_{\mathrm{dS}}(a)$.
The non-vanishing components
of the hypersurface energy-momentum tensor (\ref{eq:sab})
computed from the extrinsic curvature (\ref{eq:kab})
are interpreted as the surface energy density of the gravastar shell $\sigma$
and the isotropic surface tension $\theta$, and are given by
   \begin{eqnarray}
   \sigma & \equiv & - S_t^t = - \left[
      \frac{ \sqrt{ 1 - \mu(a) } }{ 4\pi \, a }
   \right] \; ,
   \label{eq:sigmagen} \\
   \theta & \equiv &
      - S_{\vartheta}^{\vartheta} = - S_{\varphi}^{\varphi} = - \left[
      \frac{ 1 - \mu(a) - a \mu'(a) / 2 }{ 8 \pi \, a \, \sqrt{1-\mu(a)} }
   \right]
   \label{eq:thetagen}
   \end{eqnarray}
(note that by definition surface tension has the opposite sign
of surface pressure).
The compactness functions for the dS and the RN spacetimes can be written as
   \begin{equation}
   \mu_{\mathrm{dS}}(r) = \frac{8\pi\,\rho_0}{3}\,r^2
   \quad , \quad
   \mu_{\mathrm{RN}}(r) = \frac{2M}{r} - \frac{Q^2}{r^2},
   \end{equation}
where $\rho_0$ is the constant (dark) energy density of the dS spacetime
and $M$ and $Q$ are the mass and the charge parameters of the RN spacetime.
It is convenient to use the configuration variables defined as follows,
   \begin{equation}
   \kappa \equiv |Q/M| \quad , \quad
   x \equiv \mu_{\mathrm{dS}}(a) \in (0,1) \quad , \quad
   y \equiv \mu_{\mathrm{RN}}(a) \in (0,1) \; .
   \end{equation}
(The definition of $x$ and $y$ as the compactness at the interior
and at the exterior side of the gravastar shell is compatible with
the notation used in Ref.~\cite{DHSI07}.)
For the charge to mass ratio $\kappa <1$,
the RN spacetime has the horizons at $r_{\pm}=M(1\pm\sqrt{1-\kappa^2})$
which, as $\kappa\to1$,
merge into a single (extremal) horizon of the ERN spacetime.
For $\kappa\le1$, in order to exclude the horizons from the geometry,
the gravastar surface radius must satisfy $a>r_+=M(1+\sqrt{1-\kappa^2})$.
For $\kappa>1$, the compactness is bounded from above
with its maximum $\kappa^{-2}<1$ at $r=M\kappa^2$
and there are no horizons in the RN spacetime.
However, this upper bound also forbids arbitrarily compact configurations
so the $\kappa>1$ configurations will not be further considered.
Assuming $\kappa\le 1$ for the rest of this section,
and using the configuration variables $x,y$,
the surface energy density $\sigma$ and the surface tension $\theta$
of the gravastar shell are given by
   \begin{eqnarray}
   \sigma & = & \frac{\sqrt{1-x} - \sqrt{1-y}}{4\pi\,a} \; ,
   \label{eq:sigmachgs} \\
   \theta & = & \frac{1-2x}{8\pi\,a\,\sqrt{1-x}}
   + \frac{1-\kappa^2-\sqrt{1-\kappa^2 y}}{8\pi\,a\,\kappa^2\,\sqrt{1-y}} \; .
   \label{eq:thetachgs}
   \end{eqnarray}

\subsection{Energy conditions}

We now want to determine the region in the $x,y$ configuration space
where the DEC (and therefore also the NEC and the WEC)
is satisfied for different values of $\kappa$
(as mentioned earlier we do not require the SEC to be satisfied).
The DEC requires the surface energy density to be non-negative
and the absolute value of the surface tension (or pressure)
not to exceed the value of the surface energy density, i.e.
   \begin{equation}
   0 \le |\theta| \le \sigma \; .
   \label{eq:shelldec}
   \end{equation}
The condition $\sigma\ge0$
with the surface energy density (\ref{eq:sigmachgs}) implies $x\le y$,
i.e.\ the compactness of the interior metric
must be less than the compactness of the exterior metric at the shell.
Therefore the configurations in lower right triangle of the $x,y$ plane,
shown in Fig.~{\ref{fig:one}} (left plot), are not allowed by the DEC.
Inspection of (\ref{eq:thetachgs}) reveals that $\theta\le0$
for all values of $x,y\in(0,1)$ and $\kappa\in(0,1]$,
meaning that in all configurations we are considering
the gravastar shell is under (positive) surface pressure.
The condition $|\theta|\le\sigma$ imposed by the DEC
therefore reduces to $\theta+\sigma\ge0$,
which for (\ref{eq:thetachgs})
further reduces to an inequality of the form
   \begin{equation}
   x \le f(y;\kappa) \quad , \quad y \in (0,1) \; ,
   \label{eq:fff}
   \end{equation}
where the function $f(y;\,\kappa)$ can be obtained analytically,
but is a rather complicated expression
so we discuss only its most important features.
In the case $\kappa=0$,
corresponding to the electrically neutral gravastar,
(\ref{eq:fff}) can be written as
   \begin{equation}
   x \le f(y;0) = \frac{
      60 - 36 y - 25 y^2 - (6 - 5y) \sqrt{ 100 - y ( 124 - 25 y ) }
   }{
      128 (1-y)
   } \; .
   \end{equation}
The function $f(y;0)$ has nodes at $y=0$ and $y=24/25$,
for $y\in(0,24/25)$ it satisfies $0<f(y;\,0)<y$
with a single maximum at $y=4/5$,
$f(4/5;0)=(19-\sqrt{105})/32)\simeq0.274$,
see the innermost shaded `D-shape' in Fig.~\ref{fig:one}
(configurations within the `D-shape' are allowed by the DEC).
This means that the highest surface compactness
on the exterior side of the shell of the electrically neutral gravastar
satisfying the DEC is $y = \mu_{\mathrm{RN}}(a) \to 24/25$
with $x=\mu_{\mathrm{dS}}(a) \to 0$,
i.e.\ with vanishing (dark) energy density in the interior.
This also means that, if the DEC is to be satisfied,
increasing the (dark) energy density in the interior
does not lead to higher surface compactness
of the electrically neutral gravastar.

For the charge-to-mass ratio $\kappa\in(0,1)$,
the DEC satisfying `D-shaped' region in the $x,y$ parameter space
is larger than that corresponding to the neutral configuration,
but the boundary $f(y;\kappa)$ is qualitatively similar
to the $\kappa=0$ case.
As shown in the Fig.~\ref{fig:one} (left plot),
the upper $y$-node of $f(y;\kappa)$ (the `D-shape')
for $\kappa\in(0,1)$ remains less than unity,
implying that the surface compactness
of the gravastar satisfying the DEC
is bounded from above below unity.
Again, an increase of the (dark) energy density in the gravastar interior
corresponds to a decrease of the upper bound on the surface compactness.

A qualitatively different situation is reached at $\kappa=1$.
Here we obtain
   \begin{eqnarray}
   x \le f(y;1) & = & \frac{1}{32}
      \left( \vphantom{\sqrt{\left(\sqrt{1-y}\right)}} \right.
      19 - 4 \sqrt{1-y} + 4 y \nonumber \\
   & & \mbox{} - \sqrt{ 121 + 104 \sqrt{1-y}
   - 8 y \left(15 + 4 \sqrt{1-y} - 2 y \right) }
      \left. \vphantom{\sqrt{\left(\sqrt{1-y}\right)}} \right) \; ,
   \end{eqnarray}
see Fig.~\ref{fig:one} (left plot).
The boundary of the DEC--allowed region is no longer a `D-shape'.
The upper bound on the surface compactness is no longer present,
which means that the surface compactness can reach unity arbitrarily close
without violating the DEC.
The gravitational redshift of light
emitted from the surface of such a body
could be arbitrarily high.

\subsection{Equation of state and causality}

\begin{figure}
\begin{indented} \item[]
\includegraphics{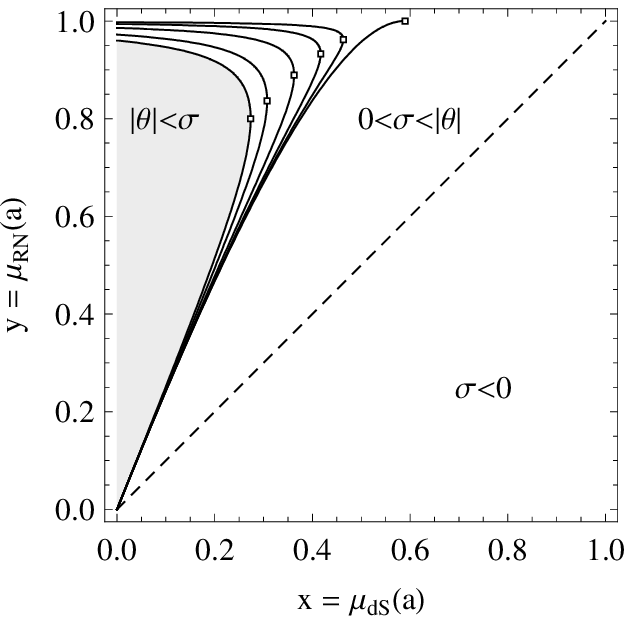} \hfil
\includegraphics{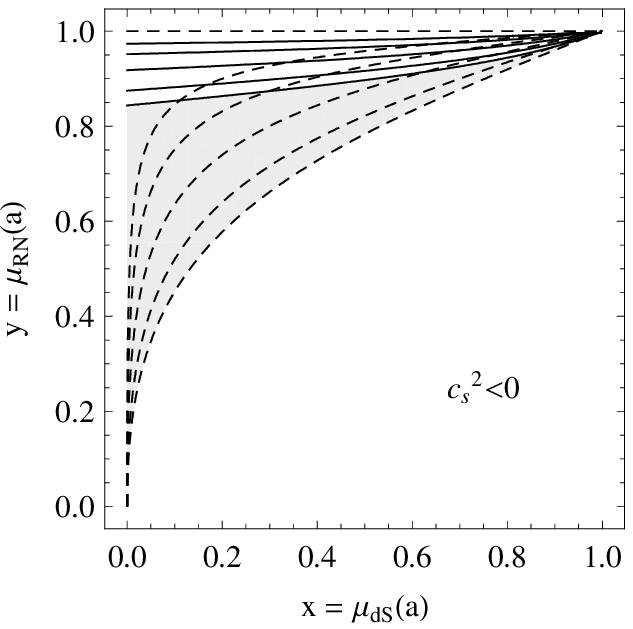}
\end{indented}
\caption{\label{fig:one}
The $\delta$-shell charged gravastar configurations shown in the $x,y$ plane.
\emph{Energy conditions (left plot):}
The configurations with $x>y$
(lower--right triangle below the dashed line)
would require negative surface energy density $\sigma$
and are thus ruled out by the DEC.
In the `D-shaped' regions bounded by solid lines
the condition $|\theta|\le\sigma$ required by the the DEC is satisfied.
Going from left to right, the solid lines correspond to
$\kappa=|Q/M|=0$ (shaded region), and subsequent lines
to $\kappa=\frac{2}{3},\frac{8}{9},\frac{26}{27},\frac{80}{81}$.
The `open top' solid line corresponds to $\kappa=1$ (ERN).
Bullets indicate highest $x$ that can be obtained with certain $\kappa$.
\emph{Causality of EOS (right plot):}
Pairs consisting of a dashed and a solid line
bound the regions in which
$c_{\mathrm{s}}^2\in[0,1]$ for different values of $\kappa=|Q/M|$.
In the regions below the dashed lines $c_{\mathrm{s}}^2<0$,
and above the solid lines $c_{\mathrm{s}}^2>1$.
The lowest pair of dashed--solid lines
bounding the shaded `bended triangle' corresponds to $\kappa=0$,
and subsequent pairs of lines correspond to to
$\kappa=\frac{2}{3},\frac{8}{9},\frac{26}{27},\frac{80}{81}$.
For $\kappa=1$ (ERN) only the dashed line is shown,
since for all configurations $c_{\mathrm{s}}^2\le0$.
}
\end{figure}

In a perfect fluid with the equation of state $p=p(\rho)$,
where $p$ is the pressure and $\rho$ is the energy density,
the quantity $c_{\mathrm{s}}^2 = \mathrm{d}p / \mathrm{d}\rho$
can be interpreted as the speed of the propagation of sound (squared).
Negative $c_{\mathrm{s}}^2$
is usually interpreted as an indication of instability,
or at least as the impossibility of propagation of sound,
while $c_{\mathrm{s}}^2>1$ indicates superluminal speed of sound
or violation of causality.
In the context of the $\delta$-shell gravastar model,
the matter comprising the shell can be understood as the
perfect fluid in 2+1 dimensions.
In Ref.\ \cite{VissGS1} a procedure has been developed
that allows one to extract the equation of state $\theta=\theta(\sigma)$.
Starting from the solutions obtained above
one can extract the equation of state $\theta=\theta(\sigma)$
from which the speed of sound (squared),
  \begin{equation}
  c_{\mathrm{s}}^2
  = - \frac{\mathrm{d}\theta}{\mathrm{d}\sigma}
  = - \left. \frac{\mathrm{d}\theta/\mathrm{d}r}{\mathrm{d}\sigma/\mathrm{d}r}
  \right|_{r=a} \; ,
  \end{equation}
becomes
  \begin{equation}
  c_{\mathrm{s}}^2 = - \frac{
      \kappa^4 \frac{\sqrt{1-y}}{1-x}
    - \frac{\sqrt{1-x}}{1-y} \left( \kappa^4 - 3 \kappa^2 + 2
        + \left( (3-y)\kappa^2 - 2 \right) \sqrt{1-\kappa^2 y} \right)
  }{
    2 \kappa^4 \sqrt{1-y} - 2 \kappa^2 \sqrt{1-x} \left(
      \kappa^2 (1-2y) + 1 - \sqrt{1-\kappa^2 y}
    \right)
  } \; .
  \end{equation}
As shown in Fig.~\ref{fig:one} (right plot),
for $\kappa\in[0,1)$ we have the $c_{\mathrm{s}}^2<0$ (non-propagating) regime
in the lower-right region of the $x,y$ square,
the $c_{\mathrm{s}}^2>1$ (superluminal) regime is in the upper-left,
while the physically acceptable region is confined in
a `bended triangular region' bounded by the 
solid curve from above ($c_{\mathrm{s}}^2=1$),
and the corresponding dashed curve from below ($c_{\mathrm{s}}^2=0$).
Starting from the lowest pair of curves (dashed and solid)
bounding the shaded $c_{\mathrm{s}}^2\in[0,1]$ (allowed)
region for the electrically neutral gravastar,
as $\kappa$ increases we see that the `bended triangle'
shifts toward higher surface compactness $y$, but also becomes narrower.
In the limit $\kappa\to1$,
the $c_{\mathrm{s}}^2\in[0,1]$ (allowed) region shrinks
so that the $c_{\mathrm{s}}^2<0$ (non-propagating) region
takes over the whole $x,y$ square.
Therefore, the $\kappa=1$ configurations (ERN),
do not satisfy the $c_{\mathrm{s}}^2\in(0,1)$ requirement.

\section{Continuous profile models \label{sec:continous}}

\subsection{The model}

Here we take the similar approach
but in the Einstein--Maxwell system governed by the equations
  \begin{equation}
  R^{\alpha}_{\beta}-\frac12 R\delta^{\alpha}_{\beta}
  = 8\pi T^{\alpha}_{\beta},
  \qquad F^{\alpha\beta}{}_{;\beta} = 4\pi j^{\alpha} \; ,
  \end{equation}
where $R^{\alpha}_{\beta}$ is the Ricci tensor, $R=R^{\alpha}_{\alpha}$,
$T^{\alpha}_{\beta} = T_{\mathrm{(fluid)}}{}^{\alpha}_{\beta}
+ T_{\mathrm{(charge)}}{}^{\alpha}_{\beta}$ is the energy-momentum tensor
of the anisotropic fluid and the electrostatic field.
The electromagnetic part of the energy-momentum tensor is given by
$T_{\mathrm{(charge)}}{}^{\alpha}_{\beta} =
\frac{1}{4\pi} \big( F^{\alpha\mu} F_{\beta\mu} - \frac14
\delta^{\alpha}_{\beta} F^{\alpha\beta} F_{\alpha\beta} \big)$,
$F^{\alpha\beta}$ is the electromagnetic field strength tensor
and $j^{\alpha}$ the source four-current.
We use the coordinates $x^{\alpha}=(t,r,\vartheta,\varphi)$ as before,
and write now the general spherically symmetric static metric simply as
  \begin{equation}
  d s^2 = g_{tt}(r) \; d t^2 + g_{rr}(r) \; d r^2 + r^2 d \Omega^2 \; .
  \end{equation}
Assuming $A_{\alpha}=\delta_{\alpha}^t A_t(r)$
for the electromagnetic four-potential,
the only non-vanishing component of the field strength tensor
$F_{\alpha\beta} = \partial_{\alpha} A_{\beta} - \partial_{\beta} A_{\alpha}$
is $F_{tr} = - F_{rt} = - \partial_r A_t$,
and the locally measured radial electric field is
$ E = \partial_r A_t / {\sqrt{-g_{tt}\,g_{rr}}} $.
The four-velocity of the static fluid element is
$u^{\alpha} = \delta^{\alpha}_t / \sqrt{-g_{tt}}$,
and the four-current of the electric charge
with density $\sigma(r)$ is
$ j^{\alpha} = \sigma \, u^{\alpha}
= \delta^{\alpha}_t \sigma / {\sqrt{-g_{tt}}} $.
The Maxwell equation reads $\left(r^2\,E(r)\right)'
=4\pi\,r^2\,\sigma/\sqrt{g_{rr}}$,
(prime denoting the derivative with respect to $r$).
Integrating from the center to some radius $r$ one obtains
  \begin{equation}
  E(r) = \frac{q(r)}{r^2} \; ,
  \qquad \mathrm{where} \qquad
  q(r) = \int_0^r 4\pi \, r'^2 \, \sigma(r') \, \sqrt{g_{rr}(r')} \, d r'
  \label{eq:qdef}
  \end{equation}
is the amount of the electric charge within the sphere of radius $r$.
The electromagnetic part of the energy momentum tensor is
  \begin{equation}
  T_{(\mathrm{charge})}{}^{\alpha}_{\beta}(r) =
  \frac{q^2(r)}{8\pi\,r^4} \, \mathrm{diag}(-,-,+,+) \; ,
  \label{eq:tem}
  \end{equation}
and the part due to the anisotropic fluid is
  \begin{equation}
  T_{(\mathrm{fluid})}{}^{\alpha}_{\beta}(r) =
  \mathrm{diag} \left( - \rho(r),
    \, p_{\parallel}(r), \, p_{\perp}(r), \, p_{\perp}(r) \right) \; ,
  \end{equation}
where $\rho$ is the energy density,
$p_{\parallel}$ is the radial and $p_{\perp}$ is the transverse pressure.
Writing the metric components $g_{tt}$ and $g_{rr}$
in terms of two functions, $m$ and $\Psi$, as
  \begin{equation}
  g_{tt}(r) = - \left( 1 - {2m(r)}/{r} \right) e^{2\Psi(r)} \quad , \quad
  g_{rr}(r) = \left( 1 - {2m(r)}/{r} \right)^{-1} \; ,
  \end{equation}
the Einstein equations give three equations:
  \begin{eqnarray}
  m'(r) & = & 4\pi \, r^2 \, \rho(r) + \frac{q(r)^2}{2r^2} \; ,
         \label{eq:einm} \\
  p_{\parallel}'(r) & = &
         - \frac{(\rho(r)+p_{\parallel}(r))(m(r)+4\pi r^3 p_{\parallel}(r)
                   - q^2(r)/2r)}{r^2(1-2m(r)/r)} \nonumber \\
         & & \mbox{} + \frac{\sigma(r) \, q(r)}{r^2\sqrt{1-2m(r)/r}}
                     + \frac2r \, (p_{\perp}(r)-p_{\parallel}(r)) \; ,
         \label{eq:einp} \\
  \Psi'(r) & = & 4\pi \, r \, \frac{\rho(r)+p_{\parallel}(r)}{1-2m(r)/r} \; .
         \label{eq:einpsi}
  \end{eqnarray}
Eq.\ (\ref{eq:einm}) defines the `mass function' $m(r)$
and (\ref{eq:einp}) is the version of the well-known TOV equation
for the anisotropic and electrically charged fluid.

The electrically neutral gravastar solutions
with continuous pressure profiles
satisfying the general requirements of Ref.~\cite{VissGS2}
were constructed in Ref.~\cite{DeBenedGS1}
from the above system (with $q=0$)
together with these essential ingredients:
\begin{itemize}
\item continuous fluid energy density profile $\rho(r)$
outwardly non-increasing and vanishing at the surface radius $r=R$,
for example
  \begin{equation}
  \rho(r) = \rho_0 \left( 1 - (r/R)^n \right) \; ; \quad n\ge 2 \; ,
  \label{eq:rhoansatz}
  \end{equation}
where $\rho_0$ is the central fluid energy density,
\item fluid pressure anisotropy Ansatz of the form
  \begin{equation}
  \frac{p_{\perp}(r) - p_{\parallel}(r)}{\rho(r)} = \frac{\alpha^2}{12}
  \left(\frac{2m(r)}{r}\right)^k
  \left(\frac{\rho(r)}{\rho(0)}\right)^l \; ; \quad k,l\ge 1 \; ,
  \label{eq:anisoansatz}
  \end{equation}
where $\alpha$ is the measure of anisotropy strength,
\item behavior of the pressures -- boundary conditions,
  \begin{equation}
  p_{\parallel}(0) = p_{\perp}(0) = - \rho(0)
  \quad , \quad
  p_{\parallel}(R) = p_{\perp}(R) = 0 \; .
  \label{eq:bcp}
  \end{equation}
\end{itemize}
For the electrically charged model,
we adopt the above ingredients,
and couple two density profiles in the following way
  \begin{equation}
  \sigma(r) = \varepsilon \, \rho(r) \, \sqrt{1-2m(r)/r} \; ,
  \label{eq:duhsigma}
  \end{equation}
with $\varepsilon$ constant \cite{Cooper-dlc78,HerreraPdL85,Gron86,PdL93}.
The equations (\ref{eq:einm}--\ref{eq:einpsi})
together with the Maxwell equation that can be written as
  \begin{equation}
  q'(r) = 4\pi \, r^2 \, \sigma(r) \, \frac1{\sqrt{1-2m(r)/r}} \; ,
  \label{eq:maxq}
  \end{equation}
close the system in four unknowns:
the two metric components $m(r)$ and $\Psi(r)$,
the fluid radial pressure $p_{\parallel}(r)$ and the charge $q(r)$.

At $r=R$ we match the interior solution
to the exterior segment of the Reissner--Nordstr\"om spacetime with the metric
  \begin{equation}
    ds^2=-\left(1-\frac{2M}{r}+\frac{Q^2}{r^2}\right)
    dt^2+\left(1-\frac{2M}{r}+\frac{Q^2}{r^2}\right)^{-1}dr^2+r^2\,
    d\Omega^2 \; ,
  \end{equation}
where  $M$ and $Q$ are the mass and the charge parameters of the RN spacetime.

The interior mass function can always be written as the sum of two functions,
  \begin{equation}
  m(r)=m_\rho(r)+m_q(r) \; ,
  \label{eq:totmass}
  \end{equation}
where
  \begin{equation}
  m_\rho(r) = \frac{q(r)}{\varepsilon} =\int_0^r 4\pi\,{r'}^2\,\rho(r')\,dr'
  \quad , \quad
  m_q(r) = \int_0^r \frac{q^2(r')}{2r'^2} \, dr' \; ,
  \label{eq:mrhomq}
  \end{equation}
subscripts suggesting the origin of contributions: fluid and charge.
The next step in the solution process is to solve the TOV (\ref{eq:einp})
for the radial pressure.
(We note here that the equations are invariant
with respect to the change of sign of the electric charge.)
Acceptable solutions of the equations,
for a given combination of $k,l,n$
in (\ref{eq:rhoansatz}) and (\ref{eq:anisoansatz})
depend on a $\rho_0$ and $\varepsilon$.
A step in cornering their values is connected
with the smooth joining of the metrics at the gravastar surface
  \begin{equation}
  m(R)  = m_{\rho}(R) + m_q(R) = M - \frac{Q^2}{2R}
  \quad \mathrm{and} \quad q(R) = \varepsilon m_{\rho}(R) = Q \; .
  \label{eq:rnjoin}
  \end{equation}
Further denoting $\kappa = |Q/M| $ as in Sec.~\ref{sec:delta},
the above conditions combine to give a quadratic in $\varepsilon$
  \begin{equation}
    \varepsilon_{1,2} = \frac1{2a\kappa}
      \left(1 \pm \sqrt{ 1 - 4 a \kappa^2 } \right)
    \quad , \qquad
      a = \frac{m_q(R)}{\varepsilon^2 m_\rho(R)} + \frac{m_\rho(R)}{2R} \; ,
   \label{eq:jconfinal}
  \end{equation}
from which we get the upper bound on the
central fluid energy density $\rho_0$.
The bound is saturated at $\varepsilon=\varepsilon_{1,2}=2\kappa$.
Two values of $\varepsilon$ lead
to the desired charge-to-mass ratio $\kappa=|Q/M|$,
one of which leads to solutions violating the DEC.

\subsection{Results}

\begin{figure}
\begin{indented} \item[] \includegraphics{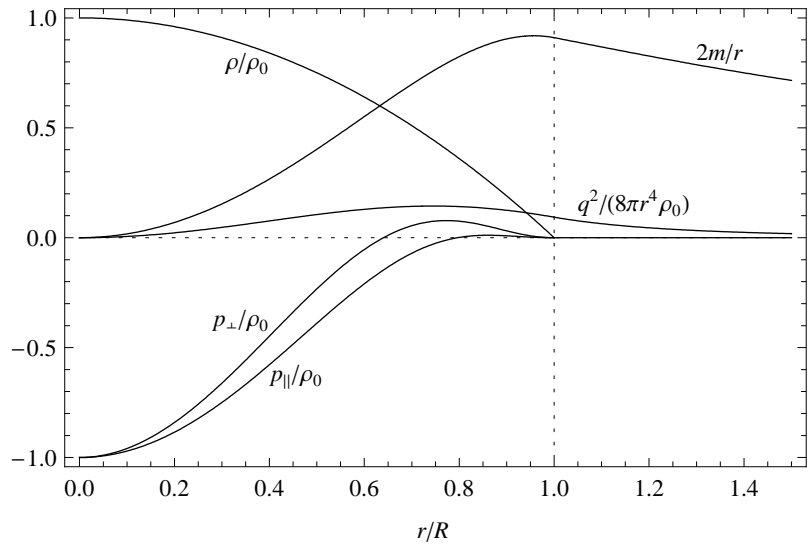} \end{indented}
\caption{\label{fig:two}
Continuous profile gravastar solution
inducing the ERN ($\kappa=|Q/M|=1$) spacetime:
energy density (\ref{eq:rhoansatz}) with $n=2$
and pressure anisotropy (\ref{eq:anisoansatz}) with $k=l=1$.
Using $\varepsilon=\varepsilon_1=\varepsilon_2=2\kappa$
gives $\rho_0 R^2=21/32\pi$, $M=Q=7R/10$,
surface compactness $\mu=91/100$,
surface redshift $Z_{\infty}=7/3$,
and pressure anisotropy strength $\alpha^2/12\simeq0.69$.}
\end{figure}

The solution displayed in Fig.~\ref{fig:two}
corresponds to the maximum allowed (fluid) density
$\rho_0=21/(32\pi R^2)$.
The presented solution corresponds to the ERN case
and it is evident that the radial and transverse pressures obey the DEC
while the compactness is safely protected from reaching unity.
The gravastar configurations obeying the energy conditions \cite{DHSI07}
show no quasi black hole behavior as described in Ref.~\cite{LemZasMimickers}.

\begin{figure}
\begin{indented} \item[]
\includegraphics{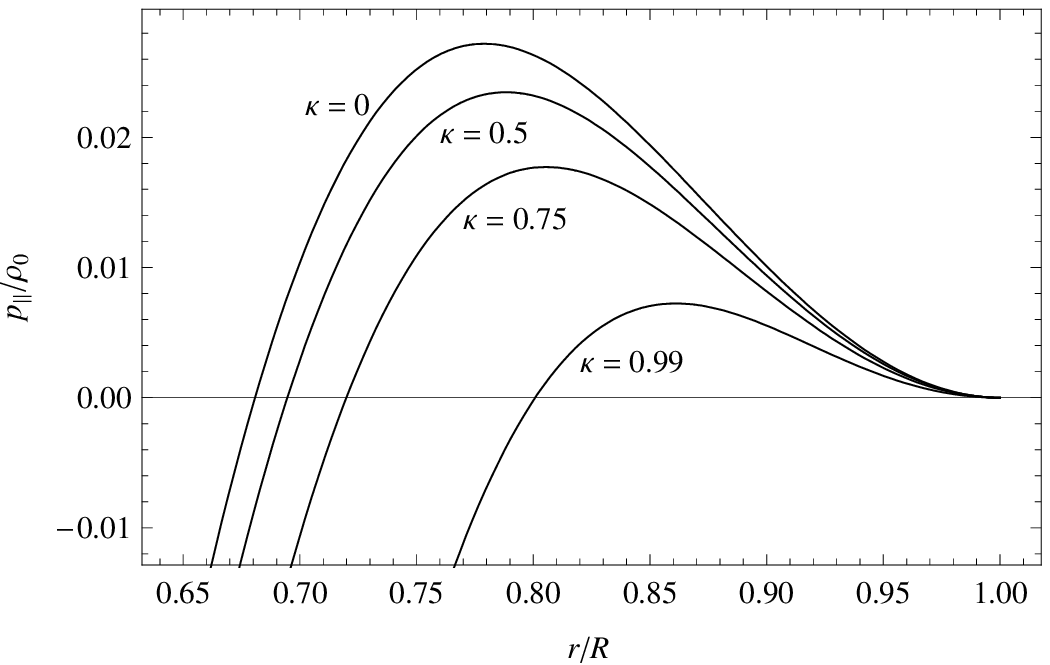} \\
\includegraphics{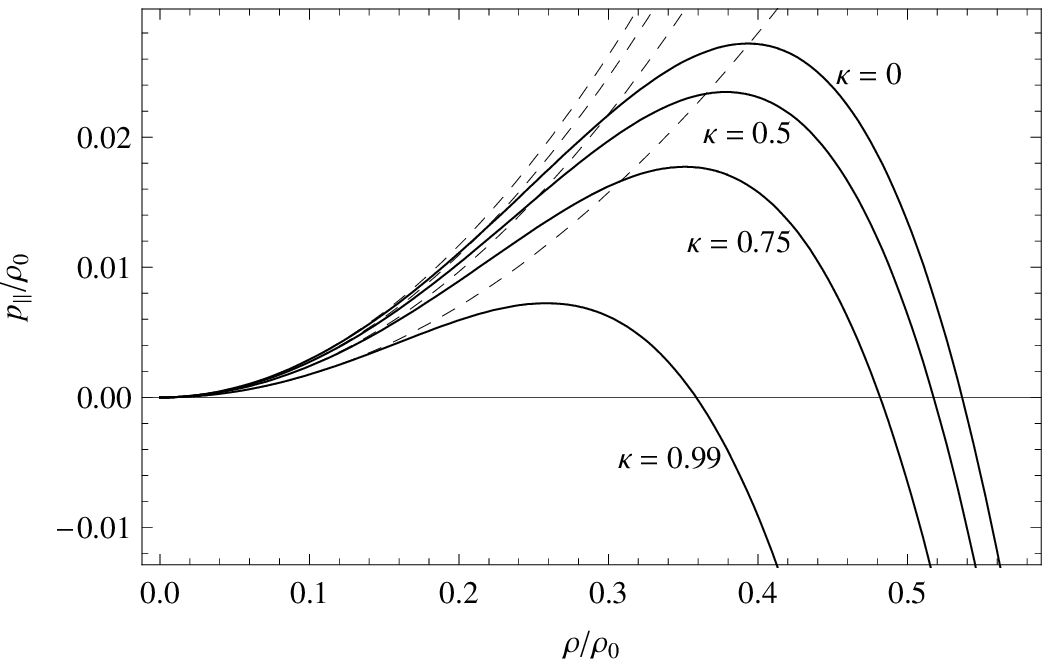}
\end{indented}
\caption{\label{fig:three}
The behavior of fluid radial pressure in the gravastar atmosphere and crust
for energy density (\ref{eq:rhoansatz}) with $n=2$
and pressure anisotropy (\ref{eq:anisoansatz}) with $k=l=1$,
for a sequence of solutions with $\rho_0 R^2 = 21/32\pi$ and
$\kappa=|Q/M|=0,0.5,0.75,0.99$:
Pressures are shown as functions of $r/R$ (upper plot)
and as functions of the fluid energy density (lower plot, solid lines).
In the polytropic equation of state
approximating the behavior of the pressures
in the outermost layer of the atmosphere, $p=k \rho^{1+1/n_p}$,
the polytropic index is $n_p\simeq1$ for all solutions
and $k$ decreases with $\kappa$ (thin dashed lines).}
\end{figure}

In the context of compact objects an important source of
information and classification criterion is the redshift,
which in the uncharged gravastar version was shown to be
within expectations for compact objects \cite{DeBenedGS1}.
The surface redshift given by
  \begin{equation}
    Z_{\infty}=-1+|g_{tt}(R)|^{-1/2} \; ,
    \label{eq:redshft}
  \end{equation}
for the perfect fluid spheres is less or equal to 2.
The anisotropy alters this value
and provides a significant increase up to 3.84
\cite{Ivanov02,Barraco03,KarmakMSM07},
while the uncharged gravastar solutions
with the continuous pressure \cite{DeBenedGS1}
give values between $1.23$ and $1.71$.
The $\delta$-shell model for the ERN case
allows the unbounded surface redshift value,
whereas the continuous density model gives the surface redshifts
that are larger than their uncharged counterparts,
e.g.\ in the ERN case shown in Fig.~\ref{fig:two}
the surface redshift is $Z_{\infty} \simeq 2.33$
with maximum compactness $\simeq 0.92$.

The results quoted so far do not depend qualitatively
on the particular choice of the profile function
for the fluid/charge density distribution:
for $n=4$, the corresponding  maximal central density
is  $\rho_0=585/(1216 \pi R^2)$,
and the maximal (internal) compactness  here is $\mu=0.93$,
so we get basically the same form of solutions.

For the general RN solutions the $r_+$ horizon is
safely positioned within the gravastar surface:
the interior compactness $\mu(r)=2m(r)/r=2(m_{\rho}(r)+m_q(r))/r$
has to be always smaller then unity.
This gives $2m(R)<R$
and since the RN mass parameter $M$ is given by (\ref{eq:totmass})
then $M\to m(R)$ for $Q\to 0$ giving $r_+\to 2M$ for $Q\to 0$.
Therefore
 \begin{equation}
  r_+ \to 2M \to 2 m(R) < R \qquad \mathrm{or} \qquad r_+ < R \; ,
  \label{eq:rplus}
  \end{equation}
and we have no ``naked horizon'' for the charged gravastar.

As a result of our approach to the charged gravastar problem
we are in the position to correlate the pressure and the fluid density.
The correlation changes from the usual matter at low densities
to that associated with de~Sitter geometry at high densities
near the center of configuration, and this change is well behaved.
It is important to note that in the atmosphere of the gravastar
(see \cite{VissGS2,DeBenedGS1})
the radial pressure has a negative gradient,
so in this region a star-like behavior is expected.
This behavior is shown in Fig.~\ref{fig:three}.
In the upper panel it is interesting to note
the change of the behavior of the radial pressure
due to the presence of charge:
by increasing the charge the fluid pressure decreases.
Since the electric energy density adds on to the fluid density,
the total mass and the compactness are increasing as well.

In the lower panel of Fig.~\ref{fig:three}
several solutions are presented all of which
show the polytropic pattern in the low density region,
  \begin{equation}
  p(\rho)=k\rho^{\gamma}=k\rho^{1+1/n_p}.
  \end{equation}
For all solutions the polytropic index is $n_p\simeq1$
and $k$ decreases with the charge-to-mass ratio $\kappa$.

The speed of sound calculated from the EOS is always subluminal
for all solutions in Figs~\ref{fig:two} and \ref{fig:three},
and is less than in the case of the uncharged gravastar.
Recall that the negative speed of sound squared
excludes the ERN case in the $\delta-$shell model.

\section{Conclusions \label{sec:concl}}

In this paper two versions
of the charged gravastar solutions have been presented.
Within the $\delta$-shell model,
solutions that satisfy the DEC have been obtained
and it has been shown that unbounded surface redshift
is possible only in the solutions inducing the ERN spacetime.
However, the analysis of the speed of sound
computed from the resulting equation of state
revealed that the ERN solutions are excluded.
In the continuous profile approach, solutions are determined
by values of the central fluid energy density $\rho_0$
and the charge-to-mass ratio $\kappa$.
Close to the surface where the pressures are non-negative
the equation of state is constructed.
It shows the polytropic behavior with the index $n_p\approx 1$.
The speed of sound is subluminal.

The obtained results help further understanding of the gravastar concept,
either as the possible alternative to the black hole
and phenomenological cosmological objects,
or simply as interesting solutions
found within the theory of General Relativity.

\section*{Acknowledgments}

We thank the referees for suggestions that helped us to
improve the paper.
We thank Andrew DeBenedictis
for reading the manuscript and for useful remarks.
This work has been supported by the Croatian Ministry of Science,
Education and Sport under the project No.\ 036-0982930-3144.

\section*{References}

\bibliography{../../relativity}

\end{document}